\documentclass[a4paper,11pt]{article}
\usepackage{pos}

\usepackage{dsfont}
\usepackage{float}
\usepackage{amsmath,bm,amsfonts}
\usepackage{color}
\usepackage{graphicx}
\usepackage[dvipsnames]{xcolor}
\usepackage{url}
\usepackage{epsfig}
\usepackage[T1]{fontenc}
\usepackage{multirow}
\usepackage{physics} 

\usepackage{booktabs} 
\usepackage{array} 
\usepackage{paralist} 
\usepackage{grffile}
\usepackage{verbatim} 
\usepackage{subfig} 
\usepackage{slashed}
\usepackage[utf8]{inputenc}
\usepackage{hyperref}
\allowdisplaybreaks[1]

\usepackage{color}
\usepackage[normalem]{ulem}

\usepackage{multirow}
\usepackage[title]{appendix}

\def\be{\begin{equation}}
\def\ee{\end{equation}}
\def\ba{\begin{eqnarray}}
\def\ea{\end{eqnarray}}

\newcommand{\slv}{\raise.15ex\hbox{$/$}\kern-.53em\hbox{$v$}}
\newcommand{\slnbar}{\raise.15ex\hbox{$/$}\kern-.53em\hbox{$\bar{n}$}}
\newcommand{\slF}{\raise.15ex\hbox{$/$}\kern-.53em\hbox{$F$}}
\newcommand{\sllbar}{\raise.15ex\hbox{$/$}\kern-.40em\hbox{$\bar{l}$}}
\newcommand{\slh}{\raise.15ex\hbox{$/$}\kern-.40em\hbox{$h$}}
\newcommand{\slP}{\raise.15ex\hbox{$/$}\kern-.53em\hbox{$P$}}
\newcommand{\slR}{\raise.15ex\hbox{$/$}\kern-.53em\hbox{$R$}}
\newcommand{\slz}{\raise.15ex\hbox{$/$}\kern-.53em\hbox{$Z$}}
\newcommand{\slzbar}{\raise.15ex\hbox{$/$}\kern-.53em\hbox{$\bar{Z}$}}
\newcommand{\slQ}{\raise.15ex\hbox{$/$}\kern-.53em\hbox{$Q$}}
\newcommand{\slK}{\raise.15ex\hbox{$/$}\kern-.53em\hbox{$K$}}
\newcommand{\slkbar}{\raise.15ex\hbox{$/$}\kern-.53em\hbox{$\bar{k}$}}
\newcommand{\slkone}{\raise.15ex\hbox{$/$}\kern-.53em\hbox{$k_1$}}
\newcommand{\slpone}{\raise.15ex\hbox{$/$}\kern-.53em\hbox{$p_1$}}
\newcommand{\slpbarone}{\raise.15ex\hbox{$/$}\kern-.53em\hbox{$\bar{p}_1$}}
\newcommand{\slptwo}{\raise.15ex\hbox{$/$}\kern-.53em\hbox{$p_2$}}
\newcommand{\slpbartwo}{\raise.15ex\hbox{$/$}\kern-.53em\hbox{$\bar{p}_2$}}
\newcommand{\slqone}{\raise.15ex\hbox{$/$}\kern-.53em\hbox{$q_1$}}
\newcommand{\slD}{\raise.15ex\hbox{$/$}\kern-.53em\hbox{$\!D$}}
\newcommand{\slC}{\raise.15ex\hbox{$/$}\kern-.53em\hbox{$C$}}
\newcommand{\slA}{\raise.15ex\hbox{$/$}\kern-.73em\hbox{$A$}}
\newcommand{\slSigma}{\raise.15ex\hbox{$/$}\kern-.53em\hbox{$\Sigma$}}
\newcommand{\slpartial}{\raise.15ex\hbox{$/$}\kern-.53em\hbox{$\partial$}}
\newcommand{\slcalP}{\raise.15ex\hbox{$/$}\kern-.63em\hbox{$\cal P$}}
\newcommand{\sleps}{\raise.15ex\hbox{$/$}\kern-.53em\hbox{$\epsilon$}}
\newcommand{\slepsbar}{\raise.15ex\hbox{$/$}\kern-.53em\hbox{$\overline{\epsilon}$}}
\newcommand{\slepsstar}{\raise.15ex\hbox{$/$}\kern-.53em\hbox{$\epsilon$}^\star}
\newcommand{\slS}{\raise.15ex\hbox{$/$}\kern-.73em\hbox{$S$}}

\newcommand{\td}{\text{d}}

\newcommand{\bk}{\mathbf{k}}

\newcommand{\bbperp}{\mathbf{b}}

\def\T{{\cal T}}
\def\td{\textrm d}
\def\tx{{\textrm x}}

\def\l{{\mathbf l}}

\title{One-loop renormalization of the quark TMD in projectile light-cone gauge}

\author[a]{Tolga Altinoluk}
\author[a]{Guillaume Beuf}
\author[b,c]{Jamal Jalilian-Marian}
\author*[a,d]{Mirja Tevio}

\affiliation[a]{Theoretical Physics Division, National Centre for Nuclear Research, Pasteura 7, Warsaw, 02-093, Poland}

\affiliation[b]{Department of Natural Sciences, Baruch College, CUNY, 17 Lexington Avenue, New York, NY 10010, USA}
\affiliation[c]{The City University of New York Graduate Center, 365 Fifth Avenue, New York, NY 10016, USA}
\affiliation[d]{
Department of Physics, University of Jyväskylä,  P.O. Box 35, 40014 University of Jyväskylä, Finland
}

\abstract{
We use the background field formalism to calculate next-to-leading-order corrections to a quark transverse momentum dependent parton distribution function (TMD) in the projectile light-cone gauge, i.e. $A^+=0$ for a left-moving target. 
We calculate the full result using the Mandelstam–Leibbrandt prescription for handling the light-cone singularity.
From the one-loop renormalization of the quark TMD, we obtain the Collins-Soper-Sterman (CSS) evolution equations. 
 
The projectile light-cone gauge is widely used in calculations within the gluon saturation regime. Our work therefore opens a way to understanding the connections between the TMD and Colour Glass Condensate (CGC) approaches within the same framework. }

\FullConference{The 33rd International Workshop on Deep Inelastic Scattering and Related Subjects (DIS2026)\\
4 - 8 May 2026\\
Bologna, Italy\\}

\begin{document}
\maketitle

\section{Introduction}

The relation between transverse-momentum-dependent (TMD) factorization \cite{Bomhof:2006dp,Bacchetta:2006tn} 
and the Color Glass Condensate (CGC) \cite{Gelis:2010nm,Albacete:2014fwa} 
is an open problem in their overlapping high-energy regime.  Evolution for TMDs is described by the Collins–Soper–Sterman (CSS) equations~\cite{Collins:1981uk, Collins:1981uw, Collins:1984kg}, which are commonly formulated in covariant or target light-cone gauges.
Whereas,  CGC calculations typically use the projectile light-cone gauge, i.e. $A^+=0$ for a left-moving hadron or nucleus. 

The correspondence between TMD and CGC factorization in their common region of validity has been studied recently for example in Refs.~\cite{Dominguez:2011wm,Marquet:2017xwy,Taels:2022tza,Caucal:2023fsf,Caucal:2024nsb,Taels:2023czt}. 
These studies focus on the back-to-back limit in dijet and dihadron production. 
In this work~\cite{Altinoluk:2026yrj}, we calculate the next-to-leading-order (NLO) corrections to the quark TMD using a set of ingredients common to both frameworks: the projectile light-cone gauge, the background field formalism, and a pure rapidity regulator~\cite{Ebert:2018gsn}. This is continuation for a similar calculation in the target light-cone gauge~\cite{Altinoluk:2025ewj}. The same formalism was also applied to obtain one-loop corrections to collinear PDFs in the earlier work~\cite{Altinoluk:2023dww}. 
Compared to the covariant gauge, the light-cone formulation of the TMD operator has a more complicated Wilson-line structure and the gluon propagator contains additional singularities which are treated with the Mandelstam–Leibbrandt (ML) prescription \cite{Mandelstam:1982cb,Leibbrandt:1983pj,Leibbrandt:1983zd}.
Understanding how TMD evolution emerges from calculations using CGC-compatible tools could pave the way toward a  common theoretical framework.

\section{Setup}

We calculate NLO corrections to a quark TMD in the projectile light-cone gauge $A^+=0$ for a left-moving target with a large momentum $P^- \gg 1 $, starting from the standard operator definition:
\begin{align}
\label{def:q_op_def_T_ord}
q^{\textrm{n.r.}}_{\textrm{unsub.}}(\tx, \mathbf{b};\mu^2) = & \lim_{Y^+\rightarrow +\infty}
\int 
\frac{\td b^{ +}}{2 \pi}    \, 
e^{-i\tx P^- b^+}
\big\langle P\big| \T \Big[ 
{\overline\Psi}(b^+, \mathbf{b}, 0^-) \, \frac{\gamma^-}{2} \, 
U(Y^+ , \mathbf{b}, 0^-;b^+, \mathbf{b}, 0^-)^{\dag}
\nonumber \\
& \times\, 
U (Y^+, \mathbf{b}, 0^- ; Y^+,0_\perp  , 0^-) 
\, 
U (Y^+, 0_\perp , 0^- ; 0)   
\Psi(0) \Big]
\big|P\big\rangle_c
\, .
\end{align}
Here the superscript ''$\textrm{n.r.}$`` denotes that, while the quark and gluon fields and coupling are renormalised, the operator definition is not yet renormalized. The subscript ''$\textrm{unsub.}$`` denotes that the rapidity divergences are not yet subtracted. The factorization above contains a future-like gauge link composed of one transverse Wilson line and two light-like Wilson lines with a small transverse separation $\bbperp$.

In the background field formalism, the quark and gluon fields are dived into non-perturbative background contribution and perturbative fluctuating fields:
\begin{align}
        \label{eq: background field definition}\Psi(x)&=\psi(x)+\delta\Psi(x)\nonumber \\
        A_\mu^a(x)&=\mathcal{A}^a_\mu(x)+\delta A_\mu^a(x)\,.
\end{align}
The first non-zero corrections to the quark TMD include two fluctuation fields and are proportional to $g^2$.
We work in the dilute regime, where background fields are small, allowing us to only keep terms that are quadratic in the background fields.

The Feynman propagator for a free gluon in the light-cone gauge and momentum space is defined as
\begin{align}
{\tilde G}_{0,F}^{\mu \nu}(k)=\,
\frac{i}{\left(k^2+i0\right)}\,
\left\{
-g^{\mu \nu}
+\frac{(k^{\mu}n^{\nu}+n^{\mu}k^{\nu})}{[n\cdot k]}
\right\}\,,
\end{align}
where $n$ is a light-like vector fixed as $n_\mu A^\mu=0$, i.e. $n^\mu=g^{\mu +}$ in the projectile light-cone gauge. We regulate the light-cone singularity at
$n\!\cdot\!k=0$ using the
ML prescription: 
\begin{align}
\frac{1}{[k^+]}=\frac{\theta(k^-)}{k^++i0}+\frac{\theta(-k^-)}{k^+-i0}\,.
\end{align}

 In the projectile light-cone gauge the rapidity divergences occur in the limit $k^-\to 0$. In order to regulate these, we insert the pure-rapidity regulator~\cite{Ebert:2018gsn} \footnote{Here, $\omega$ is a bookkeeping factor~\cite{Ebert:2018gsn} 
  which follows the relations: $\lim\limits_{\eta\to 0}\omega^2=1$ and
$\frac{\nu^+}{\nu^-}\frac{\partial}{\partial\left(\frac{\nu^+}{\nu^-}\right)}\omega^2=-\frac{\eta}{2}\,.$}:
  \begin{align}
\omega^2\bigg[
\frac{{|k^-|}}{|k^+|} \, \frac{\nu^+}{\nu^-}
\bigg]^{\eta/2} \quad\xrightarrow[]{k^2=0} \quad
\omega^2\bigg[
\frac{2 |k^-|^2}{{\bk^2}} \, \frac{\nu^+}{\nu^-}
\bigg]^{\eta/2}
\, .
\end{align}
The latter expression, for an on-shell gluon, is independent of $k^+$ and can therefore be added after integration over $k^+$.
The pure rapidity regulator allows us to express the rapidity divergences as $1/\eta$ poles in the limit $\eta\to 0$.
The rapidity regularization scale is defined as $\zeta\equiv 
2(\tx P^-)^2 \nu^+/\nu^-$.

\begin{figure}[b!]
\subfloat[Gluon emission from the quark to the lower segment of the gauge link \label{Fig:diag6}]{%
       \includegraphics[width=0.4\textwidth]{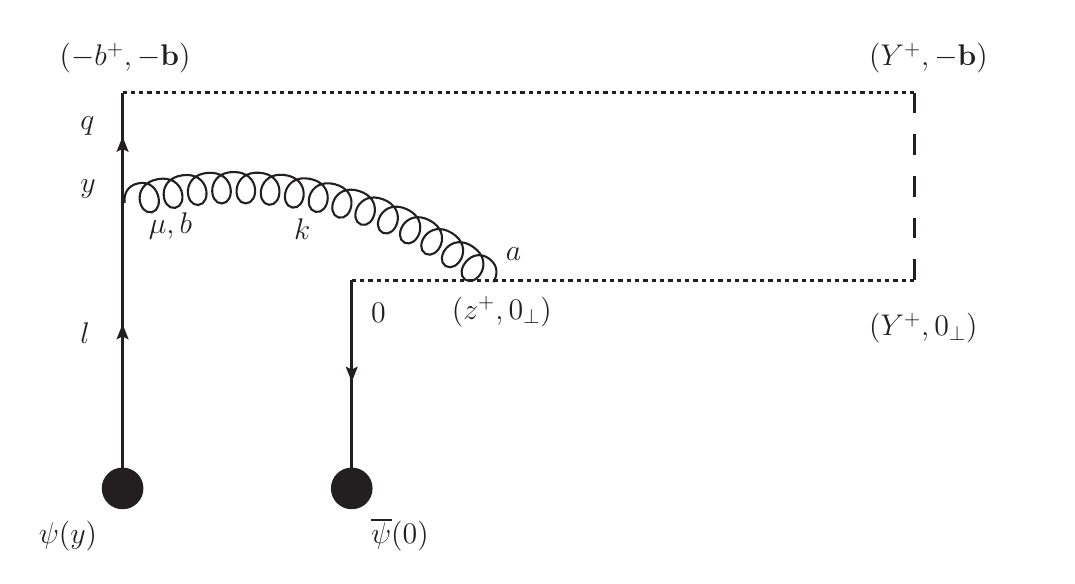}
     }
     \hfill
\subfloat[Gluon emission from the quark to the upper segment of the gauge link \label{Fig:diag4}]{%
       \includegraphics[width=0.4\textwidth]{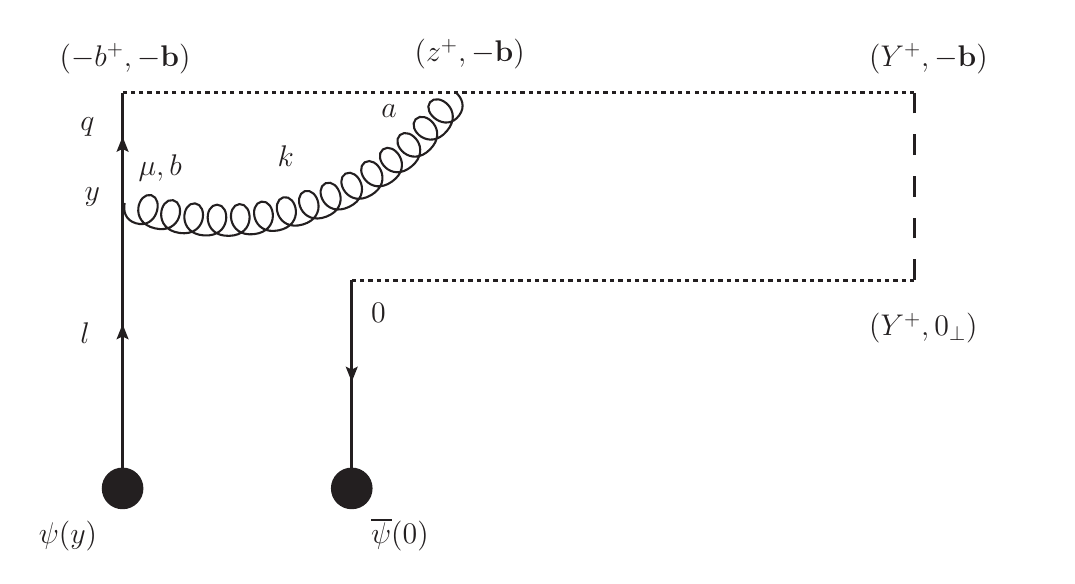}
     }
      \hfill

\subfloat[Quark-to-quark ladder diagram \label{Fig:diag1}]{%
       \includegraphics[width=0.4\textwidth]{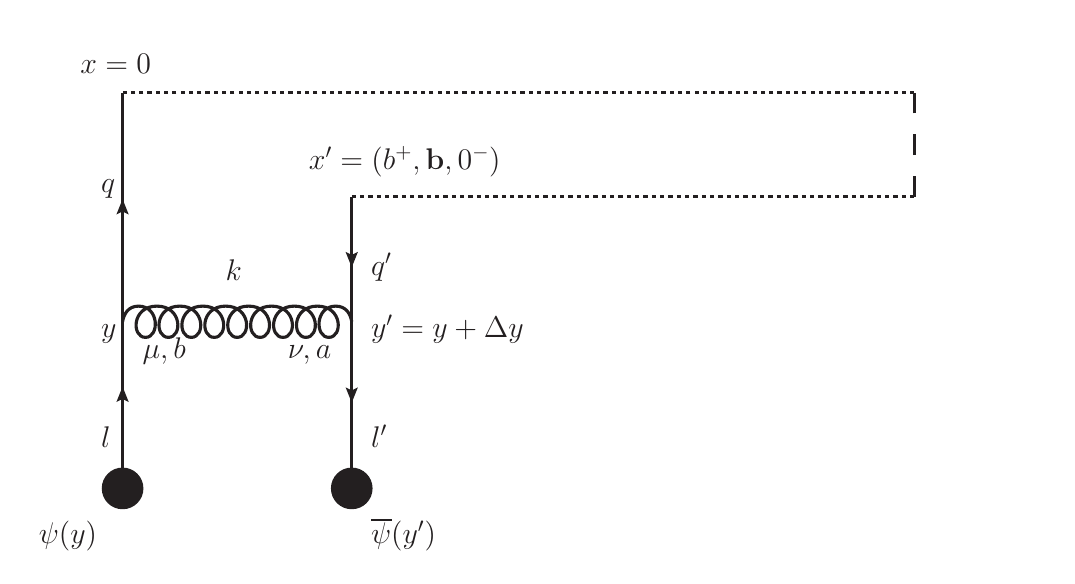}
     }
 \hfill
\subfloat[Wilson line self-energy at infinity \label{Fig:diag11}]{%
       \includegraphics[width=0.4\textwidth]{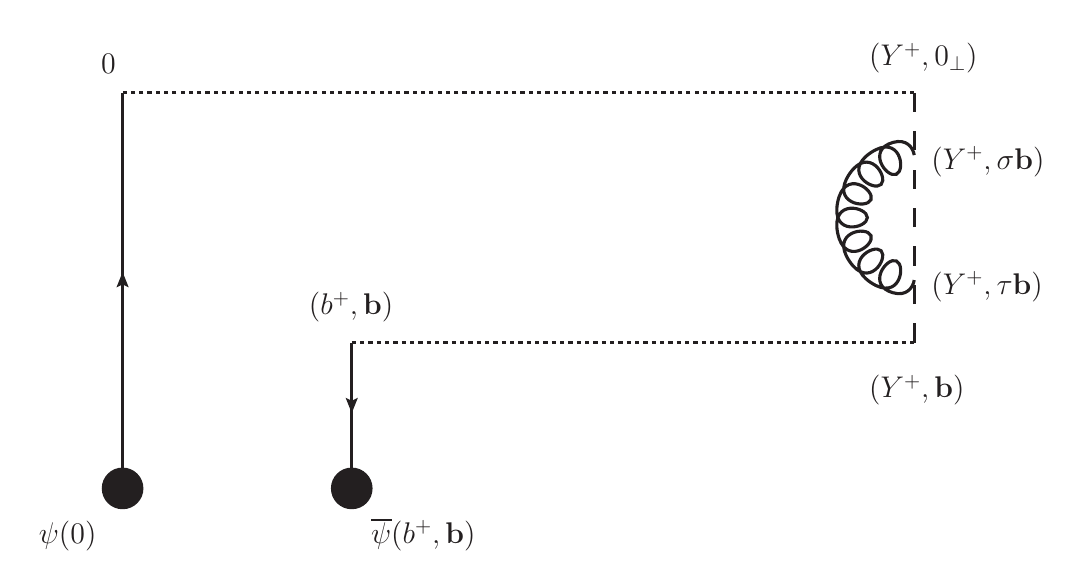}
     }

\caption{\label{Fig:real_diags 1} Diagrams giving non-vanishing contribution to NLO corrections for the quark TMD obtained by expanding around the background field in the projectile light-cone gauge.}
\end{figure}


\section{Calculation of NLO diagrams}

We substitute the quark and gluon field definitions  in the background field method, defined in Eq.~\eqref{eq: background field definition}, into the definition of the quark TMD in Eq.~\eqref{def:q_op_def_T_ord}. The NLO corrections are obtained by expanding the resulting expression and retaining only terms containing at most two fluctuation fields, which can be interpreted in terms of the quark and gluon propagators. In Figure~\ref{Fig:real_diags 1}, we present the Feynman diagrams giving non-vanishing contribution to NLO corrections to the quark TMD in the projectile light-cone gauge. Symmetric counterparts to diagrams~\ref{Fig:diag6}, \ref{Fig:diag4} are not shown but their contribution is included in the full NLO correction.

The ladder diagram~\ref{Fig:diag1} gives only a finite correction to the quark TMD at NLO and therefore does not contribute to the CSS evolution.
 Diagram~\ref{Fig:diag11}, which contains a gluon self-energy in the transverse Wilson line, contributes with a UV divergence. The calculation for this diagram is exactly the same as in the target light-cone gauge and can be found in Ref.~\cite{Altinoluk:2025ewj}. 

The combination of the rapidity and UV-divergences appear in diagrams~\ref{Fig:diag6} and \ref{Fig:diag4} and their symmetric counterparts. 
This ``double pole'' contribution to the quark TMD reads\footnote{We obtained this expression by using the Dirac equation 
 $\alpha_s\int \td ^{4-\epsilon}y e^{iy\cdot l}\langle P| \bar{\psi}(0)\gamma^- \slashed{l}\psi(y)|P \rangle \propto \mathcal{O}(\alpha_s^2)$
 for massless quarks.}
\begin{align}
&q^{\textrm{n.r.}}_{\textrm{unsub.}}(\tx, \mathbf{b};\mu^2,\zeta)\Big|_{\ref{Fig:diag6}+\ref{Fig:diag4}+\rm symmetric}^{1/\eta} \nonumber \\
&=
\frac{\alpha_s  C_F}{\pi}\,  (\pi\mu^{2} \bbperp^2)^\epsilon\,\left[\frac{\zeta \bbperp^2}{4}\right]^{\frac{\eta}{2}}\,\frac{\Gamma\left(-\frac{\eta}{2}-\epsilon\right)}{\Gamma\left(1+\frac{\eta}{2} \right)} \,\frac{\omega^2}{\eta} \,
\int \frac{\td y^+}{2\pi} e^{i \tx P^- y^+}\,
\big\langle P\big|\T\, {\overline\psi} (0)\, \frac{\gamma^-}{2}\, \psi (y^+, -\bbperp,0^-) \big|P\big\rangle_c   \nonumber \\
&= 
\frac{\alpha_s  C_F}{2\pi}\,  \Gamma (-\epsilon) (\pi\mu^{2} \bbperp^2)^\epsilon\,\left[ \frac{2\omega^2}{\eta} +\log \left(\frac{\zeta \bbperp^2}{c_0^2} \right)-\Psi(-\epsilon)+\Psi(1) \right] \, q_{\rm Bckgd}(x,\bbperp;\mu^2)
\, ,
\label{eq: double pole}
\end{align}
 where  $c_0=2e^{-\gamma_E}$ and $\Psi$ is the digamma function. Here, we used the background quark TMD:
\begin{align}
\label{def:q_Bckgd}
q_{\textrm{Bckgd}}(\mathbf{x}, \mathbf{b};\mu^2) = &
\int
\frac{d b^{ +}}{2 \pi}    \,
e^{-i\mathbf{x} P^- b^+}
\big\langle P\big| \mathcal{T} \Big[
{\overline\psi}(0) \, \frac{\gamma^-}{2} \,
\psi(-b^+, -\mathbf{b}, 0^-) \Big]
\big|P\big\rangle_c
\, .
\end{align}
From the result in Eq.~\eqref{eq: double pole}, we can obtain the expression for the rapidity renormalization factor
   \begin{align}
   Z_{\rm rap.}=1-\frac{\alpha_s C_F}{\pi}\Gamma(-\epsilon)(\pi\mu^2\bbperp^2)^\epsilon\frac{\omega^2}{\eta}+\mathcal{O}(\alpha_s^2)\,.
   \end{align}
Diagram~\ref{Fig:diag4} also gives includes a rapidity-finite but UV-divergent contribution:
\begin{align}
q^{\textrm{n.r.}}_{\textrm{unsub.}}(\tx, \mathbf{b};\mu^2)\Big|_{\ref{Fig:diag4}}^{\rm UV} = &
\frac{\alpha_s  C_F}{2\pi}\, (4\pi\mu^2)^{\epsilon}\frac{\Gamma(2-\epsilon)}{\Gamma(1-\epsilon)}\Gamma(\epsilon)\,  q_{\rm Bckgd}(x,\bbperp;\mu^2)\,.
\label{eq: UV divergent part of diag 4}
\end{align} 

The soft factor is needed to subtract double counting of soft gluons at the cross-section level~\cite{Boussarie:2023izj}.  At NLO, the soft factor is obtained from:   
\begin{align}
S(\mathbf{b}) 
&= 1 - \frac{g^2\mu^{2\epsilon}}{2}C_F
\oint_\ell dx^\mu\oint_\ell dy^\nu\,
G_{0,F,\mu\nu}(x,y) + \mathcal{O}(g^3)\,,
\label{eq:soft_factor_expanded}
\end{align}
where the integration path is a closed loop in an $(x^+,x^-)$ plane extending from $Y^+\to \infty$ to $Y^-\to \infty$ with a transverse separation $\bbperp$.
In light-cone gauge, the only non-vanishing contributions arise from the self-energies of the Wilson lines at opposite transverse parts of the gauge link.
 However, the combined result exactly cancels the UV-divergent contribution from diagram~\ref{Fig:diag11}, in Eq.~\eqref{eq: UV divergent part of diag 4}, once the soft factor is included in the quark TMD.

\section{Renormalization in Light-Cone Gauge with ML prescription}

We obtain the rapidity-subtracted, but not yet renormalized, quark TMD by including the soft factor and the rapidity renormalization factor, and finally taking the limit $\eta\to 0$:
   \begin{align}
       q^{\rm n.r.}_{\textrm{sub.}}(\tx, \mathbf{b};\mu^2,\zeta)
=  \lim\limits_{\eta\rightarrow 0}\,
Z_{\textrm{rap.}} 
\frac{1}{\sqrt{{\cal{S}}(\mathbf{b})}}
q^{\rm n.r.}_{\textrm{unsub.}}(\tx, \mathbf{b};\mu^2)\,.
   \end{align}
On the other hand, the fully renormalized quark TMD is expressed as 
 \begin{align} 
 q(x,\bbperp;\mu^2, \zeta)=Z^{UV} \, Z_2\tilde{Z}_2 \,q_{\rm sub.}^{\rm n.r}(x,\bbperp;\mu^2,\zeta)\,,
 \end{align}
     where $ Z_2\tilde{Z}_2$ in the quark field renormalization factor for the good component of the quark field spinor $\frac{\slashed{n}\slashed{\bar{n}}}{2\bar{n}\cdot n}\Psi^{(0)}$ in the projectile light-cone gauge with ML prescription. We take the existing $\overline{\rm MS }$ result from Ref.~\cite{Bassetto:1985dr}:
 \begin{align}
      Z_2\tilde{Z}_2=1-\frac{\alpha_s C_F}{4\pi}\frac{S_\epsilon}{\epsilon}\,,
 \end{align}
where $S_\epsilon=\left(4\pi e^{-\gamma_E} \right)^\epsilon$.
The UV renormalization constant can then be obtained from
 \begin{align}
 Z_2\tilde{Z}_2 q_{\rm sub.}^{\rm n.r}(x,\bbperp;\mu^2,\zeta)=&q_{\rm Bckgd}(x,\bbperp;\mu^2)  \left\{1+
\frac{\alpha_s  C_F}{2\pi}\, \left[ \frac{S_\epsilon}{\epsilon^2} +\left(\log \left(\frac{\mu^2}{\zeta} \right)+\frac{3}{2}\right)\frac{S_\epsilon}{\epsilon} \right]\right\}
\nonumber\\
&+\text{finite NLO}+\mathcal{O}(\alpha_s^2)\,,
\end{align}
resulting in
\begin{align}
     Z_{\rm UV}=1-
\frac{\alpha_s  C_F}{2\pi}\, \left[ \frac{S_\epsilon}{\epsilon^2} +\left(\log \left(\frac{\mu^2}{\zeta} \right)+\frac{3}{2}\right)\frac{S_\epsilon}{\epsilon} \right]+\mathcal{O}(\alpha_s^2)\,.
\end{align}
We notice that the renormalization constants
$Z_{\rm rap.}$ and $Z_{UV}$ remain same as in the target light-cone gauge~\cite{Altinoluk:2025ewj}.

The dependence of the renormalization scale $\mu$ and the rapidity regularization scale $\zeta$ of the fully renormalized quark TMD lie entirely on the renormalization constants $Z_{\rm rap.}$ and $Z_{UV}$. 
By writing the rapidity renormalization constant  in terms of  the bare coupling 
\begin{align}
    \alpha^{(0)}_s =\mu^{2\epsilon}\alpha_s(\mu^2) +\mathcal{O}(\alpha_s^2) \,,
\end{align}  
we obtain a $\mu$-independent expression:
     \begin{align}
     \label{eq: Zrap with bare coupling}
         Z_{\rm rap.}=1-\frac{\alpha^{(0)}_s C_F}{\pi}\Gamma(-\epsilon)(\pi\bbperp^2)^\epsilon\frac{\omega^2}{\eta}+\mathcal{O}(\alpha^{(0)2}_s)\,.
     \end{align} 
Then, the $\mu$ dependence of the renormalized quark TMD arises solely from $Z_{UV}$,
  \begin{align}
      \mu\frac{d}{d\mu}\log 
q(\tx, \mathbf{b};\mu^2,\zeta) 
=
\mu\frac{d}{d\mu} 
\log Z_{UV}\,. 
  \end{align} 
In contrast, both renormalization constants contribute to the $\zeta$ evolution:
\begin{align}
       \zeta\frac{d}{d\zeta}\log 
q(\tx, \mathbf{b};\mu^2,\zeta) 
=
\zeta\frac{d}{d\zeta} 
\log Z_{UV}  
+\zeta\frac{d}{d\zeta} 
\log Z_{\textrm{rap.}}\,.
\end{align}
Together, these results reproduce the well-known CSS evolution equations at one loop.

\section{Summary}

In this work, we calculated the NLO corrections to the unpolarized quark TMD in SIDIS using the background field formalism to obtain the perturbative corrections. The focus on this work was to understand how the CSS evolution arises from the calculation involving CGC tools such as the projectile light-cone gauge, background field formalism, and a pure rapidity regulator. The light-cone pole was regulated with the ML prescription. From the resulting rapidity and UV renormalization of the quark TMD, we recovered the CSS equations. 
A natural continuation for this work is to apply other prescriptions for the light-cone pole, as well as to calculate the NLO corrections for the gluon TMD in the same approach.

\section*{Acknowledgement}
TA is supported in part by the National Science Centre (Poland) under the research Grant No. 2023/50/E/ST2/00133
(SONATA BIS 13). GB is supported in part by the National Science Centre (Poland) under the research
Grant No. 2020/38/E/ST2/00122 (SONATA BIS 10). This material is based upon
work supported by the U.S. Department of Energy, Office of Science, Office of Nuclear Physics, within the framework
of the Saturated Glue (SURGE) Topical Theory Collaboration. JJM is supported by the US DOE Office of Nuclear Physics through Grant No. DE-SC0002307. MT is supported in part by the National Science Centre (Poland) under the research Grant No. 2024/53/B/ST2/00968
(OPUS 27).

\bibliographystyle{JHEP-2modlong.bst}
\bibliography{refs}

\providecommand{\href}[2]{#2}\begingroup\raggedright\begin{thebibliography}{10}

\bibitem{Bomhof:2006dp}
C.~J. Bomhof, P.~J. Mulders and F.~Pijlman, {\it {The Construction of gauge-links in arbitrary hard processes}},  \href{http://dx.doi.org/10.1140/epjc/s2006-02554-2}{{\em Eur. Phys. J. C} {\bf 47} (2006) 147} [\href{http://arXiv.org/abs/hep-ph/0601171}{{\tt arXiv:hep-ph/0601171}}].

\bibitem{Bacchetta:2006tn}
A.~Bacchetta, M.~Diehl, K.~Goeke, A.~Metz, P.~J. Mulders and M.~Schlegel, {\it {Semi-inclusive deep inelastic scattering at small transverse momentum}},  \href{http://dx.doi.org/10.1088/1126-6708/2007/02/093}{{\em JHEP} {\bf 02} (2007) 093} [\href{http://arXiv.org/abs/hep-ph/0611265}{{\tt arXiv:hep-ph/0611265}}].

\bibitem{Gelis:2010nm}
F.~Gelis, E.~Iancu, J.~Jalilian-Marian and R.~Venugopalan, {\it {The Color Glass Condensate}},  \href{http://dx.doi.org/10.1146/annurev.nucl.010909.083629}{{\em Ann. Rev. Nucl. Part. Sci.} {\bf 60} (2010) 463} [\href{http://arXiv.org/abs/1002.0333}{{\tt arXiv:1002.0333 [hep-ph]}}].

\bibitem{Albacete:2014fwa}
J.~L. Albacete and C.~Marquet, {\it {Gluon saturation and initial conditions for relativistic heavy ion collisions}},  \href{http://dx.doi.org/10.1016/j.ppnp.2014.01.004}{{\em Prog. Part. Nucl. Phys.} {\bf 76} (2014) 1} [\href{http://arXiv.org/abs/1401.4866}{{\tt arXiv:1401.4866 [hep-ph]}}].

\bibitem{Collins:1981uk}
J.~C. Collins and D.~E. Soper, {\it {Back-To-Back Jets in QCD}},  \href{http://dx.doi.org/10.1016/0550-3213(81)90339-4}{{\em Nucl. Phys. B} {\bf 193} (1981) 381}.
\newblock [Erratum: Nucl.Phys.B 213, 545 (1983)].

\bibitem{Collins:1981uw}
J.~C. Collins and D.~E. Soper, {\it {Parton Distribution and Decay Functions}},  \href{http://dx.doi.org/10.1016/0550-3213(82)90021-9}{{\em Nucl. Phys. B} {\bf 194} (1982) 445}.

\bibitem{Collins:1984kg}
J.~C. Collins, D.~E. Soper and G.~F. Sterman, {\it {Transverse Momentum Distribution in Drell-Yan Pair and W and Z Boson Production}},  \href{http://dx.doi.org/10.1016/0550-3213(85)90479-1}{{\em Nucl. Phys. B} {\bf 250} (1985) 199}.

\bibitem{Dominguez:2011wm}
F.~Dominguez, C.~Marquet, B.-W. Xiao and F.~Yuan, {\it {Universality of Unintegrated Gluon Distributions at small x}},  \href{http://dx.doi.org/10.1103/PhysRevD.83.105005}{{\em Phys. Rev. D} {\bf 83} (2011) 105005} [\href{http://arXiv.org/abs/1101.0715}{{\tt arXiv:1101.0715 [hep-ph]}}].

\bibitem{Marquet:2017xwy}
C.~Marquet, C.~Roiesnel and P.~Taels, {\it {Linearly polarized small-$x$ gluons in forward heavy-quark pair production}},  \href{http://dx.doi.org/10.1103/PhysRevD.97.014004}{{\em Phys. Rev. D} {\bf 97} (2018)~no.~1 014004} [\href{http://arXiv.org/abs/1710.05698}{{\tt arXiv:1710.05698 [hep-ph]}}].

\bibitem{Taels:2022tza}
P.~Taels, T.~Altinoluk, G.~Beuf and C.~Marquet, {\it {Dijet photoproduction at low x at next-to-leading order and its back-to-back limit}},  \href{http://dx.doi.org/10.1007/JHEP10(2022)184}{{\em JHEP} {\bf 10} (2022) 184} [\href{http://arXiv.org/abs/2204.11650}{{\tt arXiv:2204.11650 [hep-ph]}}].

\bibitem{Caucal:2023fsf}
P.~Caucal, F.~Salazar, B.~Schenke, T.~Stebel and R.~Venugopalan, {\it {Back-to-Back Inclusive Dijets in Deep Inelastic Scattering at Small x: Complete NLO Results and Predictions}},  \href{http://dx.doi.org/10.1103/PhysRevLett.132.081902}{{\em Phys. Rev. Lett.} {\bf 132} (2024)~no.~8 081902} [\href{http://arXiv.org/abs/2308.00022}{{\tt arXiv:2308.00022 [hep-ph]}}].

\bibitem{Caucal:2024nsb}
P.~Caucal and F.~Salazar, {\it {Dihadron correlations in small-x DIS at NLO: transverse momentum dependent fragmentation}},  \href{http://dx.doi.org/10.1007/JHEP12(2024)130}{{\em JHEP} {\bf 12} (2024) 130} [\href{http://arXiv.org/abs/2405.19404}{{\tt arXiv:2405.19404 [hep-ph]}}].

\bibitem{Taels:2023czt}
P.~Taels, {\it {Forward production of a Drell-Yan pair and a jet at small x at next-to-leading order}},  \href{http://dx.doi.org/10.1007/JHEP01(2024)005}{{\em JHEP} {\bf 01} (2024) 005} [\href{http://arXiv.org/abs/2308.02449}{{\tt arXiv:2308.02449 [hep-ph]}}].

\bibitem{Altinoluk:2026yrj}
T.~Altinoluk, G.~Beuf, J.~Jalilian-Marian and M.~Tevio, {\it {From target to projectile: CSS evolution of quark TMD in different light-cone gauges}},  \href{http://arXiv.org/abs/2606.17993}{{\tt arXiv:2606.17993 [hep-ph]}}.

\bibitem{Ebert:2018gsn}
M.~A. Ebert, I.~Moult, I.~W. Stewart, F.~J. Tackmann, G.~Vita and H.~X. Zhu, {\it {Subleading power rapidity divergences and power corrections for q$_{T}$}},  \href{http://dx.doi.org/10.1007/JHEP04(2019)123}{{\em JHEP} {\bf 04} (2019) 123} [\href{http://arXiv.org/abs/1812.08189}{{\tt arXiv:1812.08189 [hep-ph]}}].

\bibitem{Altinoluk:2025ewj}
T.~Altinoluk, G.~Beuf and J.~Jalilian-Marian, {\it {One-loop renormalization of quark TMD in the light-cone gauge: CSS evolution}},  \href{http://arXiv.org/abs/2505.20467}{{\tt arXiv:2505.20467 [hep-ph]}}.

\bibitem{Altinoluk:2023dww}
T.~Altinoluk, G.~Beuf and J.~Jalilian-Marian, {\it {Renormalization of the gluon distribution function in the background field formalism}},  \href{http://dx.doi.org/10.1103/mxtj-vblf}{{\em Phys. Rev. D} {\bf 112} (2025)~no.~3 034021} [\href{http://arXiv.org/abs/2305.11079}{{\tt arXiv:2305.11079 [hep-ph]}}].

\bibitem{Mandelstam:1982cb}
S.~Mandelstam, {\it {Light Cone Superspace and the Ultraviolet Finiteness of the N=4 Model}},  \href{http://dx.doi.org/10.1016/0550-3213(83)90179-7}{{\em Nucl. Phys. B} {\bf 213} (1983) 149}.

\bibitem{Leibbrandt:1983pj}
G.~Leibbrandt, {\it {The Light Cone Gauge in Yang-Mills Theory}},  \href{http://dx.doi.org/10.1103/PhysRevD.29.1699}{{\em Phys. Rev. D} {\bf 29} (1984) 1699}.

\bibitem{Leibbrandt:1983zd}
G.~Leibbrandt and S.-L. Nyeo, {\it {Generalized Ward Identity for the Quark Selfenergy and the Quark Quark - Gluon Vertex in the Light Cone Gauge}},  \href{http://dx.doi.org/10.1016/0370-2693(84)90783-4}{{\em Phys. Lett. B} {\bf 140} (1984) 417}.

\bibitem{Boussarie:2023izj}
R.~Boussarie {\em et.~al.}, {\it {TMD Handbook}},  \href{http://arXiv.org/abs/2304.03302}{{\tt arXiv:2304.03302 [hep-ph]}}.

\bibitem{Bassetto:1985dr}
A.~Bassetto, M.~Dalbosco and R.~Soldati, {\it {One Loop Renormalization of the {Yang-Mills} Theory With Dirac Fermions in the Light Cone Gauge}},  \href{http://dx.doi.org/10.1103/PhysRevD.33.617}{{\em Phys. Rev. D} {\bf 33} (1986) 617}.

\end{thebibliography}\endgroup

\end{document}